\documentclass[12pt]{article}
\usepackage{xcolor,pstool,enumitem,tikz,tensor,amsthm,mathtools,soul,braket,amsbsy,amsmath,amsfonts,amssymb,cite,bbold,placeins}
\usepackage{verbatim}

\usepackage[unicode,colorlinks = true,
            linkcolor = blue!70!black,
            urlcolor  = red!70!black,
            citecolor = green!55!black,
            anchorcolor = blue!70!black]{hyperref}
\usepackage[margin=0.8in]{geometry}

\setlist[enumerate]{listparindent=\parindent,parsep=0pt}

\begin{document}
\begin{titlepage}
\hfill \\
\vspace*{15mm}
\begin{center}
{\LARGE \bf A gentle introduction to \\ the cosmological multiverse}
\vspace*{15mm}

{\large Oliver Janssen}

\vspace*{8mm}

{\small
Laboratory for Theoretical Fundamental Physics \\
\'Ecole Polytechnique F\'ed\'erale de Lausanne \\
1015 Lausanne, Switzerland
}

\vspace*{1cm}
\end{center}
\begin{abstract}

\noindent \normalsize We give an introduction to the cosmological multiverse, aimed at an audience of artists. We discuss general relativity -- our modern theory of gravity -- and the \textit{cosmological constant}, which is widely believed to be responsible for the observed accelerated expansion of the universe. We then turn to a big puzzle that the cosmological constant poses, and, eventually, how the multiverse could solve this puzzle. There's no such thing as a free lunch, however: the multiverse can become arbitrarily large and old. The unsolved problem of making unambiguous predictions for observations in eternally accelerating universes is known as the measure problem of eternal inflation.

\end{abstract}

\vspace{2cm}

\textit{Adapted transcript for publication, translated to English, of a 30 minute talk given on October 3, 2024 at arts and science workshop \href{https://www.explore-unil.ch/evenement/la-mixtape-du-multivers/}{La Mixtape du Multivers} (La Grange Center, Universit\'e de Lausanne) during the  \href{https://www.explore-unil.ch/evenement/journees-de-recherche-indisciplinaires/}{Journ\'ees de recherche interdisciplinaires}}

\vspace{0.7cm}

\today

\end{titlepage}

\newpage

\tableofcontents

\section{Introduction}
My goal in these 30 minutes is to give you a basic introduction to what we mean by ``multiverse'' in physics, and how we came up with that idea. Without any ado I will start by saying what it is in one sentence: the multiverse is a hypothetical region of space and time so vast that it contains many subregions (``universes'') across which the ``laws of physics'' vary. I will try to explain what this means in more detail and why it makes sense.

I would like to stress several things at the outset:

\begin{itemize}
	\item You may have heard about another kind of multiverse, related to the ``many-worlds interpretation'' of quantum mechanics \cite{dewitt1973many}. This multiverse is more abstract and different from the more tangible cosmological multiverse we will discuss below. A few words of context: in quantum mechanics, the outcomes of most measurements are not determined. Instead, there is a set of possible outcomes, each with a certain probability that is calculable from the theory. (Compare with a biased coin that has a 39\% chance of landing on tails and a 61\% chance of landing on heads: if you flip it 1000 times, you will get roughly 390 tails and 610 heads, with an expected error on each of about 15.) The thorny question is what happens to the system upon performing a single measurement. In the many-worlds interpretation, every time a measurement occurs, the universe branches into a collection of parallel realities, in each of which a definite outcome occurs (the coin lands on heads or on tails). The frequency at which a given outcome appears across these realities corresponds to its quantum probability (39\% tails, 61\% heads). The parallel realities cannot communicate with one another. In this picture, nothing is truly lost: all possible outcomes are realized somewhere. In the more conventional Copenhagen interpretation, the universe does the opposite: it collapses into a single copy, the one whose measurement outcome is the one we actually observed. The disagreement is therefore about how to interpret probabilities in the physical theory; we use the word ``interpret'' because these various interpretations give identical predictions for all experiments and so cannot be distinguished by measurement. From the viewpoint of the scientific method, they are equivalent.
	\item The cosmological multiverse -- henceforth simply ``multiverse" -- is a hypothesis: it has not been experimentally verified.
	\item On the other hand, most of its ingredients rely on very well-tested facts about the universe (or better, our local universe), involving, in particular, gravity and quantum mechanics. That is to say, with the multiverse we are not inventing a wholly new theory of physics -- many of its components are well-established.
	\item The multiverse provides one of the very few known solutions to a deep puzzle in cosmology (the study of the evolution of the universe), namely, the cosmological constant problem \cite{Weinberg:1988cp}. We will illustrate what this is with an analogy to a game of darts. As is often the case, solutions to one puzzle give rise to more puzzles, and that is not different here: in the case of the multiverse, the main new puzzle is called the measure problem of eternal inflation. We will briefly touch upon it towards the end.
	\item If it indeed exists, it goes almost without saying that the multiverse would radically change our worldview. It would constitute a true revolution in the way we think about humanity's place in the cosmos, as radical as when we discovered the Earth was round and rotating around the Sun, or that there are many other galaxies enormously far away and that the Milky Way is not the whole thing.
\end{itemize}

\noindent Here is an outline of the ideas we'll be discussing in what follows:
\begin{enumerate}
	\item An introduction to general relativity for artists, in 3 minutes.
	\item The observation that the universe is expanding -- accelerating, in fact -- and the cosmological constant (or ``CC'' for friends), which we believe is responsible for it.
	\item The problem with the CC.
	\item A very curious calculation by \href{https://en.wikipedia.org/wiki/Steven_Weinberg}{Steven Weinberg}.
	\item Quantum tunneling, and the landscape of string theory.
	\item At this point we will have all the ingredients to introduce the multiverse, and see how it could solve the CC problem in an ``anthropic'' way.
	\item We will close with an open puzzle, which can be summarized as ``Why am I still here?''.
\end{enumerate}

\section{A brief introduction to general relativity}
To talk about the multiverse we first need to talk about gravity and a bit about cosmology. The simplest theory of gravity (massive bodies attract each other), Newton's, will not suffice for us. That theory assumes space and time are rigid, and we have learned this is not the case. Instead we must travel from the 17th to the 20th century, to Einstein, who explained that gravity is a manifestation of the curvature of space (and of time, but that is harder to imagine). That may sound abstract, but think about it this way: the effect that mass, or more generally, energy, has on space is like what a bowling ball does to a trampoline when placed on its surface. It bends the fabric of the trampoline, more so in its vicinity and less so far away from it. It is a good analogy to think of space as a fabric that can be bent. This lets one see that you don't necessarily need to have any mass to feel gravity: with or without mass, you are moving on a curved background! In particular, light -- which has no mass -- should feel the presence of gravity when moving in the vicinity of a heavy object. That is, its trajectory should be curved.

\begin{figure}[!htbp]
\centering
\includegraphics[width=225pt]{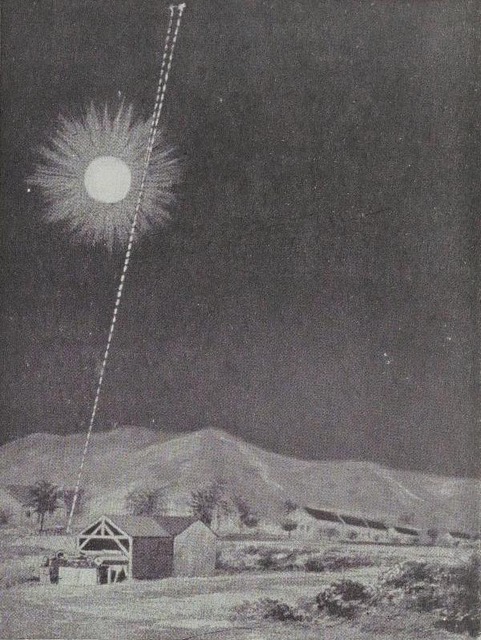}
\caption{Illustration of the bending of light by the Sun, as predicted by general relativity \cite{SobralSplendour1923}. For a modern account of the two expeditions that made the measurements during the solar eclipse of May 29, 1919, see \cite{Will:2014zpa}.}
\label{light_bending}
\end{figure}

This led to the experimental verification of Einstein's theory of gravity in 1919. During a solar eclipse, we measured that light coming from stars indeed bends around the Sun (Fig.~\ref{light_bending}). Newton's theory does not predict this.

To summarize general relativity in one sentence: it is the physical theory that follows from the idea that the fabric of spacetime is dynamical, that it can be bent and stretched by the presence of matter (like the Sun), or, more generally, by energy. Gravity \textit{is} the curvature of spacetime.

\section{The universe is expanding, and accelerating!}
Now we come to one of the most important measurements of the 20th century, done by Edwin Hubble and his team in 1929 \cite{doi:10.1073/pnas.15.3.168}: they measured that the universe is \textit{expanding}. This is perfectly consistent with Einstein's theory of gravity, since it allows space to stretch out. The \textit{expansion} of the universe means that at sufficiently large distances, all points move away from all other points.

[\texttt{Somewhat awkward practical demonstration by a theoretical physicist.}] In case you have a hard time imagining that, I will demonstrate it for you. See here a balloon, with on it several black crosses. If I blow into the balloon, then from the viewpoint of any particular cross on the balloon, all the other crosses are moving away from it. From the viewpoint of the surface of the balloon, there is no ``center'' of the expansion: it is happening everywhere in the same way. This is what it means for space to expand. It is like it is getting ``thicker'' everywhere.

\begin{figure}[!htbp]
\centering
\includegraphics[width=0.85\linewidth]{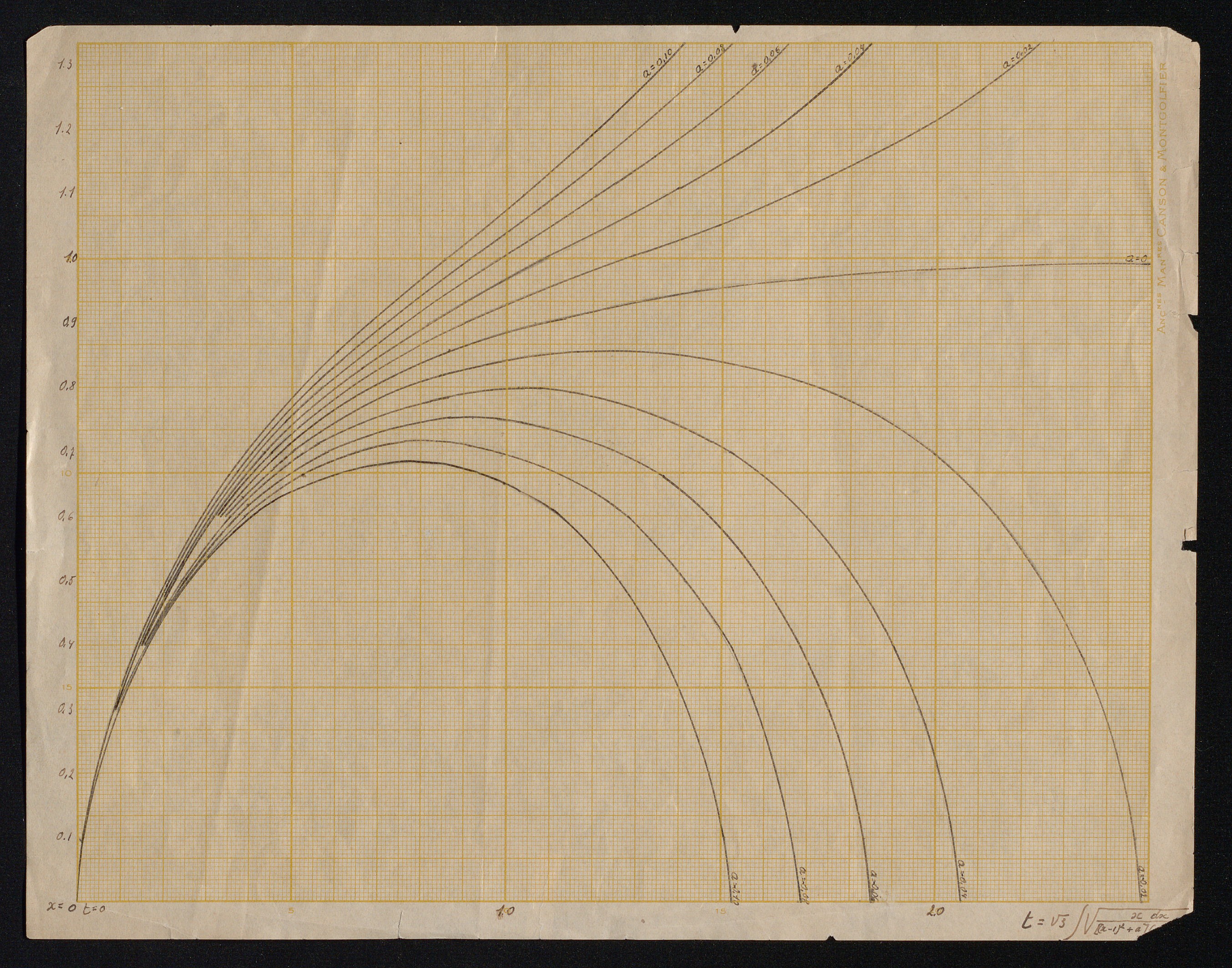}
\caption{Hand-drawn solutions to Einstein's equations by Georges Lema\^itre \cite{LemaitreArchives1927}, for a universe with positive cosmological constant (explained in \S\ref{CCsec}: this component causes the size of space, plotted on the vertical axis, to accelerate as a function of time, plotted on the horizontal axis) and positive curvature (which slows down the expansion). Depending on how much cosmological constant vs.~curvature there is, the size of space (sometimes misleadingly called the ``size of the universe'', as explained in the main text) goes through a different history in time. In 1927 Lema\^itre's graph was a theoretical speculation, but in 1998 it was confirmed that the history of our own universe looks very much like one of the top lines in this drawing. Modern data strongly suggests that our universe will keep on expanding indefinitely to the future.}
\label{lemaitre_drawing}
\end{figure}

The fact that Einstein's theory could describe an expanding (or contracting) universe was understood by a handful of people in the early 20th century, including the Belgian priest Georges Lema\^itre. In 1927 he combined Einstein's equations with the available astronomical data to derive what is now called the Hubble-Lema\^itre law, the linear relation between a galaxy's distance and its recession velocity: if a galaxy is twice as far away from us, it is moving away from us twice as fast. In Fig.~\ref{lemaitre_drawing} we show a plot from his 1927 work \cite{Lemaitre1927}, drawn by hand on millimeter paper. On the horizontal axis is time, so to go forward in time we need to move to the right. On the vertical axis is the size of space. This should not be confused with the total size of the universe, which could be finite or infinite. Instead, by ``size of space'' we mean a normalized distance between two points that are initially at rest compared to one another, and we track that distance as time goes on: it changes because the space in between the two points ``thickens'' or ``thins'' (not because the particles are moving: they remain at rest on an expanding background). One can see that the total size of the universe does not come into play here.

One of the other most important cosmic measurements of the 20th century happens near its end, in 1998 \cite{Perlmutter_1999}: we discovered that not only is the universe expanding, it is doing so in an \textit{accelerated} way! And it has been for roughly the past 5 billion years of its total age of 14 billion years. The ``stuff'' causing the universe to expand in this fashion is what physicists have called dark energy. We still do not have a clue about what dark energy is made of -- but we \textit{can} accurately model its \textit{effect} on the expansion of spacetime. One of its possible forms (and the only one we will discuss) is a cosmological constant, to which we turn next.

\section{The cosmological constant, and the problem with it} \label{CCsec}
\begin{figure}[!htbp]
\centering
\includegraphics[width=0.95\linewidth]{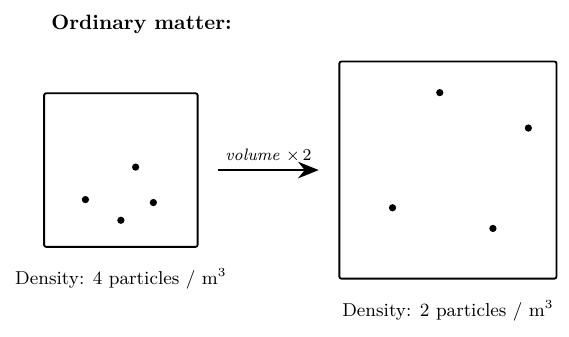}
\caption{The usual situation: we start with a fixed number of particles in a box (left), with some density; here, 4 particles per cubic meter. If we enlarge the volume of the box by two but do not add particles, the density gets halved (right).}
\label{matterFig}
\end{figure}

\begin{figure}[!htbp]
\centering
\includegraphics[width=0.95\linewidth]{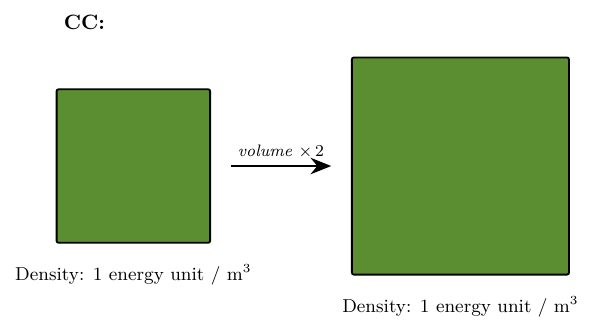}
\caption{The unusual situation of the CC: the density remains constant even though the volume increases. For the physically inclined: energy is being injected into the universe and so it appears work is being done. This is not the case because the CC comes with a \textit{negative} pressure, which exactly cancels the inflow of energy, resulting in a total work of zero.}
\label{CCFig}
\end{figure}

The cosmological constant (CC) can be understood as the tendency of ``empty'' space -- as far as we can tell -- to want to expand. We say ``constant'' because the CC has the strange property that its density does not decrease as the universe grows. Instead, its density remains constant! This is not what happens with other forms of energy that we are used to. We illustrate this property and contrast it with the situation for ordinary matter in Figs.~\ref{matterFig}-\ref{CCFig}.

Having introduced the CC, we can start to move towards the core of the multiverse idea: the ``CC problem'', a.k.a.~``the biggest problem in physics''. The ``problem'' is that the rate at which the accelerated expansion of space is happening, is \textbf{\textit{very}} small (per unit time, the increase of the expansion velocity is very small) from the viewpoint of the physics we know about (like the interactions between elementary particles). One may legitimately ask why this constitutes a ``problem''. Perhaps it is simply the nature of things, and that's that. But the situation of the CC is so striking that one cannot help thinking there must be an underlying mechanism that explains its smallness. We will try to illustrate this impression with the following analogy.

Consider throwing darts at a dartboard. We can define a ``natural length scale'' in this situation, which is the typical distance from the center where the darts would end up. If you're a professional, this will be rather small, maybe 1-2 cm. If you're not that experienced, maybe the natural length scale is 10-20cm -- in any case, let's say it is on the order of centimeters. Now imagine a friend of yours comes along and throws the dart -- as far as you can tell -- \textit{right} in the middle of the board. You come close to inspect the dart, and you think ``wow, this really is very much in the middle; I see it's in the red region of the bullseye, but this looks like it's even at the center of the bullseye itself''. You get really curious about this peculiar situation and start measuring the dart's position more and more precisely, taking a microscope to zoom in. Every time you adjust the zoom, with an accuracy of 1mm, 1$\mu$m, 1nm, etc. (1000 times smaller each time), you keep seeing that it appears \textit{right} in the middle of the bullseye (as far as you can then tell), and every time you grow more astonished that your friend managed to throw it this precisely in the middle. But eventually, you have zoomed in about 40 times -- each time having increased your precision by a factor of 1000, so that's a 1 with 120 zeroes ($10^{120}$) times zooming inwards.\footnote{Please forgive the author that this is not possible in reality.} Then, \textit{finally}, you find the dart was \textit{not} thrown exactly in the middle but that there's an unfathomably tiny, \textit{but non-zero}, displacement from the center, namely, $10^{-120}$cm. You're absolutely mystified and start to ask yourself how this is possible. Maybe your friend's dart was secretly a futuristic bullseye-seeking missile. Or maybe you got momentarily distracted when she was going to throw it, and she actually cheated by walking up to the board and placing it right in the middle using her own microscope that has an accuracy of $10^{-120}$cm (that is, 1cm divided into $10^{120}$ pieces). You're also puzzled that the dart is so close to the center, but still not actually \textit{at} the center. What has set this new, tiny length scale, which is so much smaller than the length scale of a few centimeters you expected? That's the cosmological constant problem in a game of darts.

At the time of writing, there is no agreed-upon theoretical explanation of the smallness of the cosmological constant compared to the Planck density (the natural energy density that comes out of combining general relativity with quantum mechanics; we will talk about quantum mechanics further on). The analog of the walking-to-the-board explanation would be an (unknown, and necessarily broken) symmetry in the laws of physics, while the futuristic missile explanation is analogous to a ``dynamical relaxation mechanism'', both of which are rather ad-hoc. The multiverse provides another explanation, and we will soon be ready to introduce it.

\section{Steven comes to the rescue (again)}
To summarize: from a theoretical point of view, the CC is extremely small compared to the ``natural'' energy scales that appear in known theories of physics. However, it is not zero. This is where Steven Weinberg enters our story. In 1987, \textit{before} we had measured the CC to be non-zero in 1998 (in 1987, we already knew it was small, but the CC could still have been zero, or better, unmeasurably small), he asked himself how large the CC could be, in theory, without preventing gravitationally bound structures (galaxies and the stars in them) from forming in the universe \cite{Weinberg:1987dv}. The genius of this reasoning lies in its simplicity: if the CC were too large, there would be no interesting structures, let alone \textit{life}, in the universe. In other words, with a large CC \textit{we would not be here}. Weinberg did the relevant calculation in general relativity and found that the CC cannot be much larger than the ``natural'' energy scale of quantum gravity -- the Planck density -- \textit{divided by} $10^{118}$. Several years after his seminal paper, we finally measured the CC: it is not zero! In fact, its value is not far below Weinberg's theoretical maximal value for the existence of gravitationally bound structures: how curious!

\section{Quantum tunneling and the string landscape}
So far our discussion has been about gravity. To reach the multiverse we need a second ingredient: quantum mechanics. Due to time constraints I will have to be extremely brief about this, and simply highlight one key feature of quantum mechanics that we will need: quantum tunneling. In quantum mechanics, it is possible for a particle (or any field, like the electromagnetic field, and presumably as well, the gravitational field) to ``tunnel'' through a barrier. This means that while classically it does not have enough energy to get over the barrier, there is still a small, but, crucially, \textit{non-zero} probability that the particle manages to cross the barrier.\footnote{An important example is the radioactive decay of heavy nuclei, like uranium, by the emission of a helium nucleus. For uranium-238, the decay time is about 6.5 billion years. Think of it as a helium nucleus ``trapped inside" of the uranium nucleus, constantly trying to escape from it. Classically, it would never succeed, because it does not have sufficient energy. Quantum mechanically, it will succeed in escaping approximately once every 6.5 billion years.}

\begin{figure}[!htbp]
\centering
\includegraphics[width=0.95\linewidth]{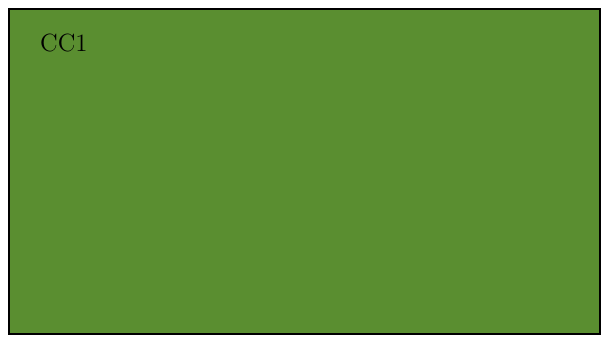}
\caption{Here we see a portion of space filled with a cosmological constant at some value 1.}
\label{CC1}
\end{figure}

\begin{figure}[!htbp]
\centering
\includegraphics[width=0.95\linewidth]{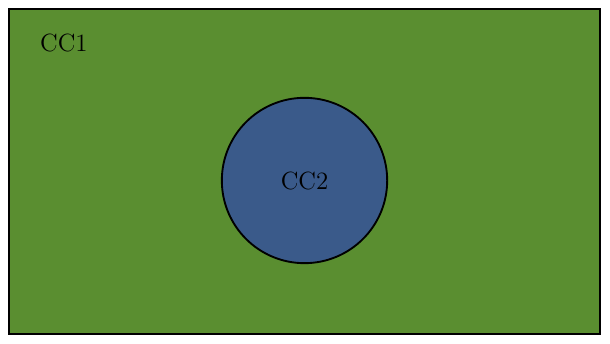}
\caption{Here a bubble has formed by quantum tunneling, filled with cosmological constant 2 -- the old CC1 still exists outside.}
\label{CC2}
\end{figure}

\begin{figure}[!htbp]
\centering
\includegraphics[width=0.95\linewidth]{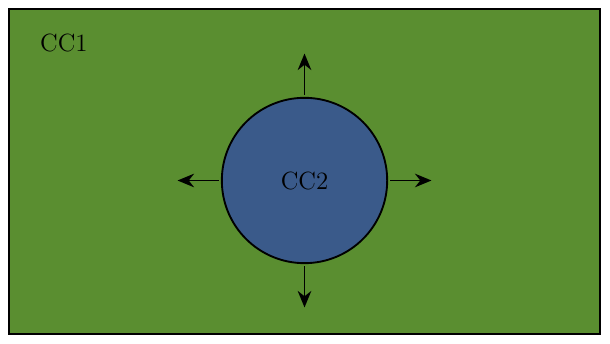}
\caption{After it has nucleated, the bubble expands into the space outside. Meanwhile this outer space itself is also expanding: everything is expanding, but at different rates determined by the local value of the CC. In the multiverse proposal, \textit{we} live in one such expanding bubble, itself located in an expanding ocean containing many other bubble universes.}
\label{CC3}
\end{figure}

\begin{figure}[!htbp]
\centering
\includegraphics[width=0.95\linewidth]{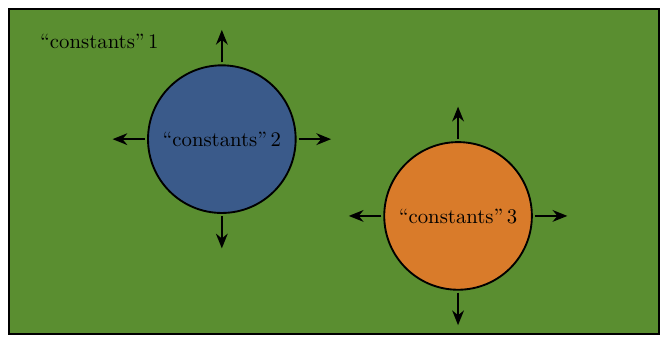}
\caption{Instead of just the CC, other constants can also vary from bubble to bubble.}
\label{CC4}
\end{figure}

The final ingredient we need to turn the multiverse into a scientific theory is string theory. One can think of it as a unified theory of gravity and quantum mechanics, the two topics we discussed so far. In this theory, what we observe around us as ``constants of nature'' may actually jump to other values by quantum tunneling. Let me stress that this kind of jumping of the constants also happens in experimentally verified theories of physics\footnote{Nota bene, string theory has not been experimentally verified!}, like the Standard Model of Particle Physics -- so this possibility is actually not as radical as it may sound. These ``constants'' include things like the masses of the elementary particles, but also the value of the CC. This ``passing to other values'' happens by the nucleation of bubbles, like in boiling water. Regions of space spontaneously come into existence by quantum tunneling (with some probability per unit time and per unit spatial volume) inside of which the constants take on new values, the old ones still existing unchanged outside of the bubble. We illustrate the idea of bubble nucleation and the jumping of the CC, and other ``constants of nature'', in Figs.~\ref{CC1}-\ref{CC4}.

\section{The multiverse}
We have finally arrived at the multiverse! It is the grand collection of all bubble-universes, each ``filled'' with their own local constants of nature, including the cosmological constant \cite{Linde:1986fd}. Each of them is expanding (or perhaps, some of them went through a phase of contraction and have vanished), and also the space in between them is expanding. To know which constants can vary, and the extent to which they can do so, we would need an overarching theory. We assume this theory is (similar to) string theory. The multiverse is therefore not a completely arbitrary theory where ``anything is possible'' -- although, to the extent that we understand string theory, the number of possibilities is enormous \cite{Bousso:2000xa}. In particular, the number of values that the CC can take in string theory appears to be gigantic. To be fair we must add a word of caution here and say that the multiverse is not a mathematically rigorous consequence of string theory\footnote{Unlike some of its other aspects, such as its consistency as a quantum theory of gravity only in 10 or 11 spacetime dimensions.}, because there is a consensus that we have only scratched the surface of what string theory really is. But within that limited understanding, I believe the evidence points to the existence of the multiverse.

In the multiverse proposal, we live in one such expanding bubble, itself located in an expanding ocean containing many other bubble-universes. This ``ocean'' may itself be a bubble-universe; one thus obtains a fractal structure of bubbles within bubbles. Now, regarding the CC problem: in this ``landscape'' of string theory, the \textit{typical} value of the CC of bubble-universes is very large (so, typically, there would be no life inside the bubble-universes of string theory). However, two factors combine to guarantee the existence of bubbles with very small CC's. On one hand, the number of possible values for the CC is immense. On the other hand, the bubble-universes expand eternally and continually allow new bubble-universes to form within them, each with a non-zero probability of having a different CC -- which leaves an infinite amount of time to realize even the most improbable configurations. There will therefore necessarily be bubbles with very small CC's, like in our local bubble-universe. So string theory provides a very large set of values the CC can take, and quantum tunneling provides the mechanism for all of these values to be realized somewhere \cite{Brown:2011ry}. In this setup, if we assume our bubble-universe is typical \textit{among the bubble-universes that contain gravitationally bound structures}, then Weinberg's calculation predicts that we should observe a value of the CC that is close to the one we have indeed measured! This is the ``anthropic'' solution to the CC problem. It is named thus because our own existence informs the answer to the calculation: we are not just asking what is the expected value of the CC in a bubble-universe, but what is this expected value in a bubble-universe that contains gravitationally bound structures. (Further conditioning on the existence of \textit{life}, for which gravitationally bound structures are presumably a prerequisite, is not crucial for this argument.)

In the dartboard analogy, the multiverse solution is one where our friend did not actually have a single attempt to throw the dart, but rather at least $10^{120}$ attempts. It would seem reasonable to assume that every time she throws the dart, it is equally likely to land anywhere on the dartboard, which is characterized by a length of order a few centimeters (or better, an area of a few $\text{cm}^2$). After $10^{120}$ throws, one would then expect that at least one dart would have ended up a distance of 1cm \textit{divided by} $10^{120}$ from the center of the bullseye, that is, $10^{-120}$cm.

To complete the argument, we must first explain why the CC (or, the distance from the dart to the center of the bullseye) is not much \textit{larger} than the value of $10^{-120}$cm we have observed: although some darts may have ended up very close to the center of the bullseye, the vast majority will not have. This is where Steven Weinberg comes in, with the striking observation that our cosmic dartboard turns out to actually be made of concrete outside of a disc of radius $10^{-118}$cm centered at the bullseye. Therefore, while our friend's darts could have hit the dartboard outside of this small disc, we would have no evidence to confirm this. In other words, \textit{given that we see a dart stuck in the board}, it must be very close to the center. In our universe, the CC must be very small, because otherwise we would not be here to observe it. This only makes sense if: 1.~there exist many values the CC could have taken, in principle; and 2.~there exists a meta-structure across which the CC varies. String theory addresses the first point, and the mechanisms of quantum tunneling and eternal expansion within bubble-universes provide the meta-structure.

Finally, we must argue why the dart did not end up much \textit{closer} to the center than $10^{-120}$cm.\footnote{To be sure: the measured value of the cosmological constant \cite{Planck:2018vyg} is $1.1 \times 10^{-52}\text{m}^{-2}$, which is associated with a length scale of order $10^{26}\text{m}$ or about 10 billion light years, which is the same order of magnitude as the size of the visible universe. The number $10^{120}$ appears as the ratio of the Planck energy density to the energy density associated with the cosmological constant.} This is simpler to explain: a disc of radius $10^{-121}$cm around the bullseye covers an area 100 times smaller than the disc of radius $10^{-120}$cm. So, given our assumption of the uniform distribution of possible dart locations, it is 100 times less likely for the dart to end up a distance $10^{-121}$cm from the bullseye than $10^{-120}$cm ($10^{-122}$cm would be 10000 times less likely, and so on).\footnote{The attentive reader will have noticed that Weinberg only showed the dartboard is made of concrete outside of a disc of radius $10^{-118}$cm, which is quite a bit larger than $10^{-120}$cm, and so, from this reasoning, a value of $10^{-120}$cm would appear pretty unlikely. We should grant Weinberg these two orders of magnitude out of 120, which might be overcome by refining the conditional probability question \cite{Tegmark:2005dy}.} We leave it to the reader's imagination to come up with other solutions to the dartboard conundrum.

\section{Boltzmann brains and the measure problem}
The multiverse solution to the CC problem is anything but minimalistic: it relies on the existence of a possibly infinite number of other universes. In addition, it completely does away with any romantic notion of ``uniqueness'' of physical theory (including its constants). This is why many physicists reject the multiverse idea. My point of view is that we have no hand in the laws of nature, and that our current understanding of quantum mechanics and gravity naturally leads to the multiverse, which is absolutely bewildering indeed.

The multiverse may solve one big problem, but it introduces another one along with it: the mechanism we described above allows the multiverse to become arbitrarily large and old. In such a scenario, counter-intuitive things start happening very late in the evolution history of the bubble-universes, which eternally expand in an accelerated fashion, because of quantum mechanics \cite{Dyson:2002pf}. Apart from the nucleation of bubbles, other objects can also spontaneously nucleate into existence. The probability of that happening is very small, but, crucially, non-zero. After a certain (long) amount of time, it will surely happen.

After a bubble-universe has expanded for a very long time -- all the stars have burned out, and all the black holes have evaporated -- it will eventually become empty. However, continued accelerated expansion has the curious feature that it endows otherwise empty space with a small, non-zero temperature. It is this tiny but non-zero temperature that allows things to spontaneously nucleate out of the ``vacuum'' (empty space): eventually, a ``brain'' would spontaneously come into existence (with high probability, it would also spontaneously turn back into the vacuum from whence it came in a very short time). This brain could be configured to have all of anyone's memories in it, believing it lives in a large world full of people, and that it just ate a sandwich. Now, at a given very late time in the universe, there will be many of these ``freak observers'', a.k.a.~Boltzmann brains. And at a later time after that, there will be even more, because the universe has expanded by some amount in the meantime. If we look at the entire history of such a universe, there is us, the observers that live close to the beginning of the expansion and that are presumably not vacuum fluctuations.\footnote{This is plausible because we observe a long, causal sequence of events -- something that would be exceedingly unlikely for a Boltzmann brain. But in the end it is an assumption.} But then there are the very large (or even infinite) number of freak observers that appear much later on. In the most straightforward way of counting, shouldn't \textit{we} be one of those freak observers living in the asymptotic future of our local bubble-universe? If one would pick a random observer among all the observers in a given bubble, this is what we would conclude, because there are only a few observers of the ``normal'' type\footnote{A mere 120 billion or so have existed so far, as far as we know!} while there are infinitely many Boltzmann brains. This problem of infinities, and how to make sensible predictions in the multiverse, is called the measure problem of eternal inflation \cite{Freivogel:2011eg} (inflation in cosmology means accelerated expansion). An excellent modern account of these topics and much more, aimed at the interested layperson, can be found in \cite{hertog2023origin}.

\section*{Acknowledgements}
We thank Romain Bionda and Aur\'elien Maignant for extensive comments on earlier versions of this article.

\phantomsection
\addcontentsline{toc}{section}{\refname}
\bibliographystyle{klebphys2}
\bibliography{refs}
\end{document}